\documentclass[
 prd,preprint,longbibliography,
 showpacs,showkeys,lengthcheck,
 nofootinbib,tightenlines,onecolumn,notitlepage,
 preprintnumbers,superscriptaddress
]{revtex4-1}

\usepackage[utf8]{inputenc}
\usepackage{newtxtext,newtxmath}
\usepackage[mathcal]{euscript}

\usepackage{graphicx}
\usepackage[dvipsnames]{xcolor}
\graphicspath{{figures/}}
\usepackage{tikz-feynman}
\usetikzlibrary{decorations.pathreplacing,calligraphy}
\usepackage{hyperref}
\hypersetup{colorlinks=true,citecolor=blue,linkcolor=blue,urlcolor=blue}
\usepackage{orcidlink}

\usepackage{diagbox}
\usepackage{booktabs}
\usepackage{siunitx}
\AtBeginDocument{%
  \heavyrulewidth=.08em
  \lightrulewidth=.05em
  \cmidrulewidth=.03em
  \belowrulesep=.65ex
  \belowbottomsep=0pt
  \aboverulesep=.4ex
  \abovetopsep=0pt
  \cmidrulesep=\doublerulesep
  \cmidrulekern=.5em
  \defaultaddspace=.5em
}

\usepackage{amsmath}
\usepackage{slashed}
\usepackage{bm}
\usepackage{bbm}
\usepackage{mathtools}
\usepackage{tensor}

\usepackage[math]{cellspace}
\usepackage[draft,commentmarkup=uwave]{changes} 
\definechangesauthor[color=green!50!black]{Adam}
\definechangesauthor[color=blue!70!black]{Jerry}

\def\la{\langle}\def\ra{\rangle}

\usepackage{soul} 

\def\s{\sigma}

\def\m{\mu}
\def\n{\nu}

\def\t{\tau}

\newcommand{\eq}[1]{Eq.~(\ref{#1})}

\def\D{\Delta}

\def\b{\beta}

\def\n_p{n_\parallel}

\begin{document}

\title{
  Isospin  violation and  the neutron weak magnetic moment 
}

\author{Gerald A. Miller \orcidlink{0000-0003-2443-3639\,}}
\email{miller@uw.edu}
\affiliation{Department of Physics, University of Washington, Seattle, WA 98195-1560, USA}

\begin{abstract}
   The recent measurement of weak magnetism in neutron $\b$ decay ~\cite{Camarasa-Domene:2026jtx} obtained a value that is consistent with the conserved vector current hypothesis requiring the assumption of isospin conservation. The influence of the quark-mass difference and electromagnetic  effects that violate this conservation law are examined using non-relativistic constituent quark models and estimated  to be of the order of at most $1 \%$.
 \end{abstract}

\preprint{NT@UW-26-26}

\maketitle

 Ref.~\cite{Camarasa-Domene:2026jtx} recently  combined   $\beta$ decay asymmetry measurements with
the neutron lifetime and the $\mathcal{F}t$ value of nuclear superallowed $0^+\to0^+$ decays to 
make the first observation of weak magnetism (WM) in neutron decay. 

The existence of WM is a consequence of
 the conserved vector current  (CVC) hypothesis ~\cite{Gershtein:1955fb,Feynman:1958ty},   a cornerstone of the standard model. It relates the weak vector current to the isovector electromagnetic current as members of the same isospin triplet. The immediate result is that  the weak-vector form factors  are directly related to  isovector magnetic moments. 
This means that   CVC  predicts that weak magnetism exists~\cite{Gell-Mann:1958sai}. Just as the matrix element of the electromagnetic current features a Pauli term ($\sigma^{\mu\nu} q_\nu$) representing the anomalous magnetic coupling, the weak vector current generates an identical   structure whose form factor is directly related by CVC to electromagnetic observables. 
 WM has  been observed in some nuclear $\beta$ transitions,  with varying dependence  on nuclear-structure input~\cite{Severijns:2006dr}.

The simplest $\beta$-decay process is that of the free neutron, but 
weak magnetism had never been directly measured prior to Ref.~\cite{Camarasa-Domene:2026jtx}.   WM is the  {only} non-negligible form factor beyond the leading $V\!-\!A$ coupling  contributing to the differential decay rate at first order in recoil~\cite{Jackson1957,Holstein1974,Hayen2018,CaoZhou2023,Ivanov2021}, so it is important for precision studies.  

The definition of the WM form factor,$g_M$ appears in 
   the matrix element of the weak current, 
  decomposed as:
\begin{align}
  \la p|V^\m_+|n\ra=  \bar u_p\left[ \frac{p_n^\m+p_p^\m}{2\bar M} +\frac{g_M}{2\bar M}i\s^{\mu\nu} q_\nu\right]u_n
\label{defV}\end{align}
where $u_{n,p}$ is the neutron (proton) spinor, $q =p_p-p_n$
is the momentum transfer, and $\bar M=(M_n+M_p)/2$ is
the average nucleon mass. Here, $g_M$  denotes
the form factor at $q^2 = 0.$ The $q^2$ dependence 
is  safely neglected in neutron beta decay. The relevant component of the vector current, expressed in terms of quark operators, is 
\begin{align}
    V_+^\mu=\bar u\,\gamma^\mu d.
\end{align}
The matrix element of the electromagnetic current is given by
\begin{align}
    J^\mu=\,\langle\bar N|\bar q\gamma^\mu\left(\frac16 +\frac12 \tau_3\right)q|N\rangle=\bar u_N\left( \frac{p_N+p_N'}{2 M_N}F^N_1(q^2) +\frac{\s^{\mu\nu}}{2 M_N}i q_\nu G_M^N(q^2)\right),
\label{defJ}
\end{align}
where $N$  represents either the proton (p) or neutron (n).  The operator $\t_3$ accounts for the vector $V$ term,

Under CVC,  the WM coupling $g_M$
is linked to the magnetic moments of the proton and
neutron, yielding ~\cite{Gell-Mann:1958sai}:
\begin{align}
  g_M=\m_p-\m_n  =4.71
  \label{gcvc}
\end{align} 
in units of nucleon magneton.
Ref.~\cite{Camarasa-Domene:2026jtx}  found that  $g_M = 4.47(87)$, in agreement with the CVC prediction. 

The CVC expression requires the assumption of isospin conservation, a symmetry that is only approximate. Isospin violation is caused by up-down quark mass difference and electromagnetic effects~\cite{Miller:1990iz,Miller:2006tv}. 
Such effects are expected to be of order $\frac{M_n-M_p}{\bar M}\approx 10^{-3}$ or 
$\frac{m_d-m_u}{(m_d+m_u)/2}\approx 10^{-2}$, very small compared to the current experimental uncertainty of 0.87.
Nonetheless, the newfound ability to measure $g_M$ merits an assessment of the size of the effects of isospin violation. This, using  non-relativistic quark models~\cite{Isgur:1979ed,Isgur:1979be,Close:1979ux},  is presented here.

Given \eq{defV} and \eq{defJ} the next step is to evaluate the matrix elements. The first step is to realize that the neutron and proton and down and up quarks are related by a particular rotation   in isospin
space.   I  briefly describe the formalism, elaborated in several
reviews~\cite{Miller:1990iz,Miller:2006tv}.  The starting point is to use the approximate
invariance of the Lagrangian under the interchange of $u$ and $d$
quarks. This makes it worthwhile to define the charge symmetry (cs)
operator, which is an isospin  rotation by 180$^\circ$ about the
y-axis
(taking the z-axis to be associated with the charge). This is defined
by
\begin{equation}
d=  P_{\rm cs}u,  
\end{equation}
with 
$P_{cs} = \exp(i\pi T_2),$
and
$T_2=\frac12{q}^\dagger \tau_2 q,$
where $q$ is the light (u,d) quark field operator.
 The Hamiltonian consists of a charge symmetry conserving term $H_0$
and a 
breaking term $H_1$  such that 
\begin{equation}
H = H_0 + H_1,
\end{equation}
with 
$ 
[H_0, P_{\rm cs}] = 0.
$
The unperturbed states are denoted by a subscript 0 and have a mass $\bar M$ that is the average of the proton and neutron masses.
It is sufficient to  work to first order in  $H_1$ such that the physical proton is
expressed in terms of the unperturbed states 
by 
\begin{equation} 
\mid p \rangle = \mid p \rangle_0 + \frac{1}{\bar M-H_0} \Lambda H_1 \mid
p\rangle_0,
\label{wf}
\end{equation}
with a similar equation for the neutron.
The quantity $\Lambda$ is a projection operator to states
orthogonal to the unperturbed ground state isospin doublet.
The unperturbed neutron and proton state vectors are related by 
 charge symmetry:
\begin{equation}
\mid n \rangle_0 = P_{\rm cs} \mid p \rangle_0.
\label{n}\end{equation}

A set of non-relativistic, constituent-quark models  described in Refs.~\cite{Miller:1997ya,Crawford:2022yhi} is used as a tool 
to a first estimate of the effects of cs  breaking.     With these models, the necessary
evaluations are not difficult  and one gains insight into the physics
of charge symmetry breaking.

To proceed further,  the Hamiltonian, $H_0$,  and its
ground state wave function must be specified. Oscillator confinement is used. Thus,
\begin{equation}
 \mid p \rangle_0 = \mid \Psi\rangle
\frac{1}{\sqrt{2}} \left(\mid \phi_ S \rangle
\mid \chi_S\rangle + \mid \phi_A\rangle  \mid \chi_A
\rangle\right)\label{su6}
\end{equation}
where SU(6) spin-isospin wave functions are used, with spin wave functions
for a spin up proton given by 
\begin{align}
    |\phi_S\ra&=\frac{1}{\sqrt{6}}|\left((\uparrow\downarrow+\downarrow\uparrow)\uparrow-2\uparrow\uparrow\downarrow\right)\ra\\
|\phi_A\ra&=\frac{1}{\sqrt{2}}|\left((\uparrow\downarrow-\downarrow\uparrow)\uparrow \right)\ra.
\end{align}
For the proton, the $\chi_{S,A}$ are obtained by replacing $(\uparrow,\downarrow)$  with $(u,d)$. The bare neutron wave function is obtained by using  \eq{n}. The spatial part of the wave function is given by 
\begin{equation}
\langle \vec{r}_i\mid \Psi\rangle=\Psi(\rho,\lambda)=
N e^{-(\rho^2 + \lambda^2) / 2 \beta}. \label{ho}
\end{equation}
Here $\vec{\rho}\equiv {1\over \sqrt{2}}(\vec{r}_1-\vec{r}_2)$ and 
$\vec{\lambda}\equiv {1\over \sqrt{ 6}}(\vec{r}_1 +\vec{r}_2 -2\vec{r}_3)$ and $N$ is a normalization constant.

 In this model the isovector part of the nucleon magnetic moment is given by
 \begin{align}
     \m_N=M_N\la N|\frac12 \sum_i \t_3(i)\frac{\sigma_3(i)}{m_i}|N\ra,
\label{em} \end{align}
with $|N\ra$ representing either the proton or neutron and $m_i$ denoting the up or down quark mass: $m_i=\bar m+\frac12 \Delta m_i \t_3,$ $\D m=m_u-m_d<0,\,\bar m=\frac12(m_d+m_u).$
The weak magnetism term is similarly obtained from the matrix element:
\begin{align}
  g_M=\bar M\la p| \sum_i \frac{\sigma_3(i) \tau_+(i)}{m(i)}|n \ra.
\label{weak}\end{align}

The effects of $H_1$ on the nucleon wave functions of \eq{wf} are assessed first. The charge symmetry breaking (CSB) terms of $H_1$ due to the down-up quark mass difference, its effects on the kinetic energy operator, one-gluon exchange potential and the electromagnetic interaction are detailed in Refs.~\cite{Miller:1997ya,Crawford:2022yhi}. These yield no contribution.
 This is because the operators $\sum_i \s_3(i)\t_\mu(i)$ excite the $\D$ baryon, but the  CSB terms of $H_1$   do not convert the nucleon to  a $\D$. Thus, the entire effects of CSB are caused by the explicit effects of the nucleon and quark mass differences that enter in \eq{em} and \eq{weak}.

 Given the SU(6) wave functions, the evaluations of the matrix elements leads to the results:
 \begin{align}
     &\m_p=\frac{\bar M}{\bar m} +\frac{\bar M}{\bar m} \frac{\D M}{\bar 2M} -\frac{\bar M}{\bar m}\frac{\D m}{\bar  m} \frac{7}{54}\\
     &\m_n=-\frac23\frac{\bar M}{\bar m} +\frac23\frac{\bar M}{\bar m} \frac{\D M}{2\bar M} -\frac{\bar M}{\bar m}\frac{\D m}{\bar  m} \frac{1}{27}\\
     &g_M= \frac53 \frac{\bar M}{\bar m}(1+\frac{\D m}{2\bar  m})
 \end{align}
When the mass differences are neglected, we see that $g_M=\m_p-\m_n.$

The fractional change in the value of $\m_p-\m_n$ caused by CSB effects is $\D_\m^{\rm CSB}=\frac{\D M}{5\bar M}-\frac{\D m}{18 \bar m}$,
and the fractional change in the value of $g_M$ caused by CSB effects is
$\D_{g_M}^{\rm CSB}=\frac{\D m}{2\bar m}$.  The nucleon masses are known to a very high precision~\cite{Mohr:2024kco}. The average constituent quark mass is 337 MeV, a value that gives the proton magnetic moment to be the measured value of 2.79, if CSB effects are ignored. The model-dependent values of quark mass differences range from -4.5 to -6.9 MeV~\cite{Crawford:2022yhi}.
This gives a range of values of $\D_m^{\rm CSB}$ from -0.0016 to -0.0020 and a range of values of $\D_{g_M}^{\rm CSB}$ from 0.0067 to 0.010.
The CSB effects on the magnetic moments are of  order of -$10^{-3}$ and those on weak magnetism are of  order 10$^{-2}$.
These are much smaller than what any conceivable measurement could achieve.
\bibliographystyle{unsrt}
\bibliography{main}

@article{Isgur:1979ed,
    author = "Isgur, Nathan",
    title = "{Isospin violating mass differences and mixing angles: the role of quark masses}",
    reportNumber = "Print-79-0759 (TORONTO)",
    doi = "10.1103/PhysRevD.21.779",
    journal = "Phys. Rev. D",
    volume = "21",
    pages = "779",
    year = "1980",
    note = "[Erratum: Phys.Rev.D 23, 817 (1981)]"
}

@book{Close:1979ux,
    author = "Close, F. E.",
    title = "{An Introduction to Quarks and Partons}",
    isbn = "978-0-12-175152-4",
    publisher = "Academic Press",
    address = "London",
    year = "1979"
}

@article{Isgur:1979be,
    author = "Isgur, Nathan and Karl, G.",
    title = "{Ground State Baryons in a Quark Model with Hyperfine Interactions}",
    doi = "10.1103/PhysRevD.20.1191",
    journal = "Phys. Rev. D",
    volume = "20",
    pages = "1191--1194",
    year = "1979"
}

@article{Mohr:2024kco,
    author = "Mohr, Peter J. and Newell, David B. and Taylor, Barry N. and Tiesinga, Eite",
    title = "{CODATA recommended values of the fundamental physical constants: 2022*}",
    eprint = "2409.03787",
    archivePrefix = "arXiv",
    primaryClass = "hep-ph",
    doi = "10.1103/RevModPhys.97.025002",
    journal = "Rev. Mod. Phys.",
    volume = "97",
    number = "2",
    pages = "025002",
    year = "2025"
}

@article{Miller:1997ya,
    author = "Miller, Gerald A.",
    title = "{Nucleon charge symmetry breaking and parity violating electron proton scattering}",
    eprint = "nucl-th/9711036",
    archivePrefix = "arXiv",
    reportNumber = "DOE-ER-41014-40-N97",
    doi = "10.1103/PhysRevC.57.1492",
    journal = "Phys. Rev. C",
    volume = "57",
    pages = "1492--1505",
    year = "1998"
}

@article{Crawford:2022yhi,
    author = "Crawford, Jacob W. and Miller, Gerald A.",
    title = "{Charge-symmetry-breaking effects on neutron {\ensuremath{\beta}} decay in nonrelativistic quark models}",
    eprint = "2209.10603",
    archivePrefix = "arXiv",
    primaryClass = "nucl-th",
    reportNumber = "nt@uw-22-13",
    doi = "10.1103/PhysRevC.106.065502",
    journal = "Phys. Rev. C",
    volume = "106",
    number = "6",
    pages = "065502",
    year = "2022"
}

@article{Miller:1990iz,
    author = "Miller, G. A. and Nefkens, B. M. K. and Slaus, I.",
    title = "{Charge symmetry, quarks and mesons}",
    doi = "10.1016/0370-1573(90)90102-8",
    journal = "Phys. Rept.",
    volume = "194",
    pages = "1--116",
    year = "1990"
}

@article{Miller:2006tv,
    author = "Miller, Gerald A. and Opper, Allena K. and Stephenson, Edward J.",
    title = "{Charge symmetry breaking and QCD}",
    eprint = "nucl-ex/0602021",
    archivePrefix = "arXiv",
    reportNumber = "NT-UW-06-05",
    doi = "10.1146/annurev.nucl.56.080805.140446",
    journal = "Ann. Rev. Nucl. Part. Sci.",
    volume = "56",
    pages = "253--292",
    year = "2006"
}

@article{Jackson1957,
  author  = {Jackson, J. D. and Treiman, S. B. and Wyld, H. W.},
  title   = {Possible Tests of Time Reversal Invariance in Beta Decay},
  journal = {Physical Review},
  volume  = {106},
  number  = {3},
  pages   = {517--521},
  year    = {1957},
  doi     = {10.1103/PhysRev.106.517}
}

@article{Holstein1974,
  author  = {Holstein, Barry R.},
  title   = {Recoil Effects in Allowed Beta Decay: The Elementary Particle Approach},
  journal = {Reviews of Modern Physics},
  volume  = {46},
  number  = {4},
  pages   = {789--814},
  year    = {1974},
  doi     = {10.1103/RevModPhys.46.789}
}

@article{Hayen2018,
  author  = {Hayen, Leendert and Severijns, Nathal and Bodek, Kazimierz and Mougeot, Xavier and Rozpedzik, Dagmara},
  title   = {High Precision Analytical Description of the Allowed Beta Spectrum Shape},
  journal = {Reviews of Modern Physics},
  volume  = {90},
  number  = {1},
  pages   = {015008},
  year    = {2018},
  doi     = {10.1103/RevModPhys.90.015008},
  eprint  = {1709.07530},
  archivePrefix = {arXiv},
  primaryClass = {nucl-th}
}

@article{CaoZhou2023,
  author  = {Cao, Hui-Yun and Zhou, Hai-Qing},
  title   = {Recoil Contributions in Neutron Beta Decay and Their Corrections to the Correlation Coefficients},
  journal = {arXiv preprint arXiv:2302.04647},
  year    = {2023},
  eprint  = {2302.04647},
  archivePrefix = {arXiv},
  primaryClass = {nucl-th},
  doi     = {10.48550/arXiv.2302.04647}
}

@article{Ivanov2021,
  author  = {Ivanov, A. N. and Höllwieser, R. and Troitskaya, N. I. and Wellenzohn, M. and Berdnikov, Ya. A.},
  title   = {Corrections of Order $ O(E_e^2/m_N^2)$ , Caused by Weak Magnetism and Proton Recoil, to the Neutron Lifetime and Correlation Coefficients of the Neutron Beta Decay},
  journal = {arXiv preprint arXiv:2010.14336},
  year    = {2021},
  eprint  = {2010.14336},
  archivePrefix = {arXiv},
  primaryClass = {hep-ph},
  doi     = {10.48550/arXiv.2010.14336}
}

@article{Camarasa-Domene:2026jtx,
    author = "Camarasa-Domene, Miriam and Gonz{\'a}lez-Alonso, Mart{\'\i}n and Novoa-Brunet, Mart{\'\i}n",
    title = "{First observation of weak magnetism in neutron decay}",
    eprint = "2610.26792",
    archivePrefix = "arXiv",
    primaryClass = "hep-ph",
    month = "10",
    year = "2026"
}

@article{Gershtein:1955fb,
    author = "Gershtein, S. S. and Zeldovich, Ya. B.",
    editor = "Trutnev, Yu. A.",
    title = "{Meson Corrections in the Theory of Beta Decay}",
    journal = "Zh. Eksp. Teor. Fiz.",
    volume = "29",
    pages = "698--699",
    year = "1955"
}

@article{Severijns:2006dr,
    author = "Severijns, Nathal and Beck, Marcus and Naviliat-Cuncic, Oscar",
    title = "{Tests of the standard electroweak model in beta decay}",
    eprint = "nucl-ex/0605029",
    archivePrefix = "arXiv",
    doi = "10.1103/RevModPhys.78.991",
    journal = "Rev. Mod. Phys.",
    volume = "78",
    pages = "991--1040",
    year = "2006"
}

@article{Feynman:1958ty,
    author = "Feynman, R. P. and Gell-Mann, Murray",
    editor = "Brown, L. M.",
    title = "{Theory of Fermi interaction}",
    doi = "10.1103/PhysRev.109.193",
    journal = "Phys. Rev.",
    volume = "109",
    pages = "193--198",
    year = "1958"
}

@article{Gell-Mann:1958sai,
    author = "Gell-Mann, Murray",
    title = "{Test of the Nature of the Vector Interaction in beta Decay}",
    doi = "10.1103/PhysRev.111.362",
    journal = "Phys. Rev.",
    volume = "111",
    pages = "362--365",
    year = "1958"
}

\end{document}